\documentclass[preprint,NumberedRefs]{JASA}

\usepackage{siunitx}
\usepackage{float}
\usepackage{dblfloatfix} 
\usepackage{graphicx}
\usepackage{subcaption}
\usepackage{mathtools}
\usepackage{romannum}
\usepackage{todonotes}

\newcommand{\ima}{\mathrm{j}}

\DeclareMathOperator{\e}{e}

\begin{document}

\title[]{A data-driven two-microphone method for in-situ sound absorption measurements}



\author{Leon Emmerich}
\affiliation{Chair of Vibro-Acoustics of Vehicles and Machines, Department of Engineering Physics and Computation, TUM School of Engineering and Design, Technical University of Munich, Boltzmannstrasse 15, 85748 Garching, Germany}

\author{Patrik Aste}
\affiliation{The Marcus Wallenberg Laboratory for Sound and Vibration Research, Department of Engineering Mechanics, KTH Royal Institute of Technology, Teknikringen 8, 10044 Stockholm, Sweden}

\author{Eric Brand\~ao}
\affiliation{Acoustical Engineering Program \& Civil Engineering Graduate Program, Federal University of Santa Maria (UFSM), Santa Maria-RS, 97050-140, Brazil}

\author{Mélanie Nolan}
\affiliation{Acoustic Technology, Department of Electrical and Photonics Engineering, Technical University of Denmark, DK-2800, Kgs. Lyngby, Denmark}

\author{Jacques Cuenca}
\affiliation{Siemens Industry Software NV, Interleuvenlaan 68, 3001 Leuven, Belgium}

\author{U. Peter Svensson}
\affiliation{Department of Electronic Systems, Norwegian University of Science and Technology, Trondheim, Norway}

\author{Marcus Maeder}
\affiliation{Chair of Vibro-Acoustics of Vehicles and Machines, Department of Engineering Physics and Computation, TUM School of Engineering and Design, Technical University of Munich, Boltzmannstrasse 15, 85748 Garching, Germany}

\author{Steffen Marburg}
\affiliation{Chair of Vibro-Acoustics of Vehicles and Machines, Department of Engineering Physics and Computation, TUM School of Engineering and Design, Technical University of Munich, Boltzmannstrasse 15, 85748 Garching, Germany}

\author{Elias Zea}
\altaffiliation{Corresponding author: zea@kth.se}
\affiliation{The Marcus Wallenberg Laboratory for Sound and Vibration Research, Department of Engineering Mechanics, KTH Royal Institute of Technology, Teknikringen 8, 10044 Stockholm, Sweden}

\email{zea@kth.se}


\preprint{}		

\date{\today} 

\begin{abstract}
This work presents a data-driven approach to estimating the sound absorption coefficient of an infinite porous slab using a neural network and a two-microphone measurement on a finite porous sample. 
A 1D-convolutional network predicts the sound absorption coefficient from the complex-valued transfer function between the sound pressure measured at the two microphone positions. 
The network is trained and validated with numerical data generated by a boundary element model using the Delany-Bazley-Miki model, demonstrating accurate predictions for various numerical samples. The method is experimentally validated with baffled rectangular samples of a fibrous material, where sample size and source height are varied. The results show that the neural network offers the possibility to reliably predict the \textit{in-situ} sound absorption of a porous material using the traditional two-microphone method as if the sample were infinite. The normal-incidence sound absorption coefficient obtained by the network compares well with that obtained theoretically and in an impedance tube. The proposed method has promising perspectives for estimating the sound absorption coefficient of acoustic materials after installation and in realistic operational conditions. 

\end{abstract}


\maketitle

\section{Introduction}

Determining the sound absorption behavior of acoustic materials in their intended application is of great interest in various engineering contexts. In contrast to impedance tube measurements~\citep{DIN1, DIN2} or measurements in a reverberation chamber~\citep{DIN3}, \textit{in situ}  and free-field measurement methods make it possible to obtain angle-dependent absorption data.  These methods generally rely on an analytical model of the sound field above the sample~\citep{brandao2015review}, allowing to determine its sound absorption properties from simple measurements of the sound pressure in its vicinity~\citep{takahashi2005situ,hald2019situ,alkmim2021angle,li1997use,mommertz1995angle}. Existing models to describe the acoustic field can be divided according to their wave propagation assumption into plane-wave models and spherical-wave models. Although the plane-wave approximation holds for large source-sample distances or high frequencies, the spherical wave assumption also applies to lower frequencies~\citep{brandao2015review,li1997use}.A comprehensive review of \textit{in~situ} methods up to 2015 can be found in~\citep{brandao2015review} and several more recent methods have been proposed in~\citep{richard2017estimation,nolan2020estimation,alkmim2021angle,dupont2020characterization,dupont2022characterization, Eser23, JonM23}. 

Despite their attractiveness compared to standardized methods, the underlying models used to retrieve the sound absorption properties generally rely on the assumption that the sample of interest has infinite dimensions. This assumption is not verified in actual measurements of finite-sized samples, leading to uncertainties in the inferred absorption properties. Typically, these uncertainties introduce oscillatory artifacts in the sound absorption coefficient at low frequencies~\citep{de1973mathematical,kimura2002required,brandao2012estimation}. This so-called edge effect or edge-diffraction effect is especially noticeable for increasing wave incidence angles and decreasing sample sizes~\citep{de1973mathematical,thomasson1980absorption,brandao2012estimation,hirosawa2009comparison,otsuru2009ensemble,kimura2002required,hirosawa2009investigation}. Additionally, the edge diffraction is larger for greater source distances in relation to the sample and increasing distance between the sensors and the sample~\citep{brandao2012estimation}. 

In recent years, considerable effort has been devoted to modeling and mitigating edge diffraction effects.
Brand\~{a}o et al.~\citep{brandao2012estimation} simulated the edge-diffraction effect of finite-sized samples using a Boundary Element Method (BEM). The edge diffraction resulted in a measured absorption coefficient that oscillates around the absorption coefficient of the corresponding infinite sample. The simulated sound absorption coefficients affected by edge-diffraction effects showed close accordance to the sound absorption calculated from actual measurements. Ottink~et~al.~\citep{ottink2016situ} proposed an experimental microphone-array method combined with a sound field model accounting for the finiteness of the sample.
The approach yielded fairly good results for wave incidence angles up to $85^\circ$. Finally, Brand\~{a}o and Fernandez-Grande~\citep{brandao2022analysis} numerically investigated the wavenumber spectrum above finite samples to characterize the edge effect and reconstruct the surface impedance.

Following a different approach, several studies have made use of machine-learning approaches to investigate the sound absorption behavior of sound absorptive materials~\citep{bianco2019machine,jeon2020estimation,gardner2003neural,liang2022estimation,JohD24,eser4452034hybrid,Eser2023}. Eser et al.~\citep{eser4452034hybrid} showed that it is possible to predict the frequency-dependent and complex-valued surface impedance using 1D-convolutional neural networks while maintaining the frequency resolution. However, the method is based on impedance tube measurements and does not allow a direct investigation of a material's angle-dependent sound absorption behavior. Zea et al.~\citep{zea2023sound} presented a deep-learning approach to predict the angle-dependent sound absorption coefficient at $14$ octave-band frequencies from absolute sound pressure spectra measurements at $12 \times 12$ microphone positions above the absorber. A 2D residual neural network was trained and validated using numerical data generated by a BEM and the Delany-Bazley-Miki (DBM) model~\citep{miki1990acoustical,zea2023dataset}. The approach showed reasonable generalization abilities for various sample parameters and numerical setups, but its experimental success was limited~\cite{Aste2024}. M\"uller-Giebeler et al.~\citep{muellergiebeler2024} proposed a hybrid physics- and data-driven model to inversely estimate the Johnson-Champoux-Allard-Lafarge (JCAL) parameters of finite porous materials above a rigid floor. The main idea was to fit the sound pressure measured above the sample with the sum of an analytical specular sound field and the edge-diffracted sound field produced by a neural network, both of these fields being functions of the set of JCAL parameters. The approach was experimentally validated with various kinds of wool and PU foam of various square sizes. While both methods in~\citep{zea2023sound} and~\citep{muellergiebeler2024} can produce angle-dependent absorption coefficients, the methods differ in (i) their material models (DBM vs. JCAL), (ii) experimental effort (microphone array vs. single microphones; baffled vs. unbaffled samples), and (iii) target function (sound absorption coefficient spectrum vs. scalar values of JCAL parameters). 

This work proposes a deep-learning approach to predict the free-field absorption coefficient of finite-sized samples using the well-recognized and comparatively simple two-microphone method~\citep{allard1985measurements,minten1988absorption}. In this way, the proposed method enables the determination of the sound absorption behavior of the material with a widely used technique.
The methodology entails a 1D residual neural network trained to predict the angle-dependent sound absorption coefficient between \SI{100}{\hertz} and \SI{2000}{\hertz} from measurements of the complex-valued transfer function between the spectra of the two microphones. The network is trained and validated on BEM simulations of porous absorbers similar to those in Zea et al.~\citep{zea2023dataset}. Furthermore, the network is experimentally tested with two fibrous samples of different sizes, varying incidence angles, and source positions. The absorption coefficients predicted by the network are compared with those obtained with impedance tube measurements at normal incidence and analytical predictions from the DBM model. 

The paper is organized as follows. Section~\ref{Sec. 2} summarizes the two-microphone method and the boundary element method. Section~\ref{Sec. 3} explains the numerical data generation and the sound absorption measurements and introduces the proposed neural network. Section~\ref{Sec. 4} evaluates and discusses the performance of the proposed method using numerical data and measurements on porous absorbers. 

\section{Theory}
\label{Sec. 2}

\subsection{Two-microphone method}
\label{Sec. 2b}

\begin{figure}[H]
\includegraphics[width=0.5\textwidth]{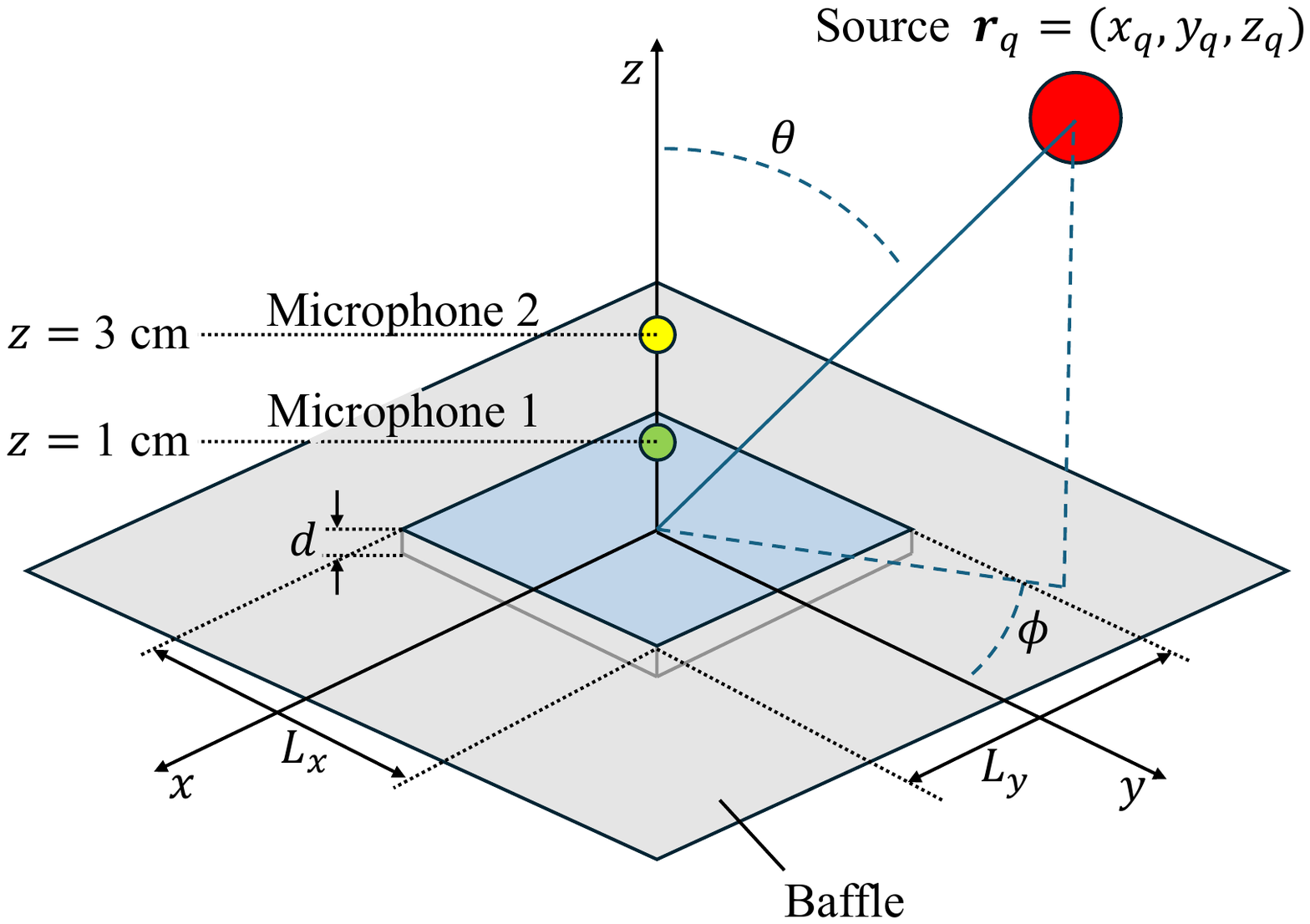}
\centering
\caption{Schematic of the two-microphone method above a baffled porous layer, flush-mounted with a rigid backing.}
\label{fig: figure_01}
\end{figure}

The two-microphone method proposed by Allard et al.~\citep{allard1985measurements} is an experimental procedure to determine the \textit{in situ} sound absorption properties, such as the sound absorption coefficient and the surface impedance. A schematic visualization of the experimental setup can be seen in Fig.~\ref{fig: figure_01}. 

A finite-sized sample of dimensions $L_{x} \times L_{y}$ and thickness $d$ is placed over a rigid layer and flush-mounted in a rigid baffle. Two microphones are positioned along the normal above the sample at positions $ z_{1}$ and $z_{2}$. The microphone positioning is a trade-off between a sufficiently close placement to the sample's surface to minimize edge effects and a required inter-microphone distance to capture useful lower frequency information~\citep{brandao2015review,allard1985measurements}. A monopole sound source is located at $\textbf{r}_{q}=[x_q,y_q,z_q]^\mathrm{T}$.
From the measured sound pressure spectra at the two microphone positions, $p_{1}(f)$ and $p_{2}(f)$, the transfer function at frequency $f$ is defined as~\citep{allard1989situ}
\begin{equation}
    H_{12}(f) = \frac{p_{1}(f)}{p_{2}(f)}.
    \label{eq. transfer function}
\end{equation}
In an experimental setup, the transfer function is corrected by dividing $H_{12}$ by a microphone calibration transfer function $H_{c}$, which is measured above a rigid surface and with the microphone membranes facing each other with a distance of less than \SI{1}{\milli\meter}. In this way, inherent phase differences between the microphones are compensated for~\citep{brandao2015review,champoux1988Hc}. 
Assuming specular reflection of spherical waves from the image source at $\textbf{r}'_{q} = [x_{q}, y_{q}, -z_{q}]^{\mathrm{T}}$, the reflection coefficient can be obtained as~\citep{li1997use}
\begin{equation}
    R(f) = \dfrac{\dfrac{\e^{-\ima k_{0}||\textbf{r}_{1}-\textbf{r}_{q}||}}{||\textbf{r}_{1}-\textbf{r}_{q}||} - H_{12} \dfrac{\e^{-\ima k_{0}||\textbf{r}_{2}-\textbf{r}_{q}||}}{||\textbf{r}_{2}-\textbf{r}_{q}||}}{H_{12} \dfrac{\e^{-\ima k_{0}||\textbf{r}_{2}-\textbf{r}'_{q}||}}{||\textbf{r}_{2}-\textbf{r}'_{q}||} - \dfrac{\e^{-\ima k_{0}||\textbf{r}_{1}-\textbf{r}'_{q}||}}{||\textbf{r}_{1}-\textbf{r}'_{q}||}},
    \label{eq. R pwa}
\end{equation}
where the microphone positions $\textbf{r}_{1}=[0,0,z_1]^{\mathrm{T}}$ and $\textbf{r}_{2}=[0,0,z_2]^{\mathrm{T}}$, $k_{0}=2\pi f/c_{0}$ is the wave-number in air, $c_{0}$ the speed of sound in air,
$\ima$ is the imaginary unit, and $||\cdot||=\sqrt{\sum |\cdot|^{2}}$ denotes the $\ell_{2}$-norm of a vector. With this, the sound absorption coefficient is then 
\begin{equation}
    \alpha(f) = 1 - |R(f)|^{2}.
    \label{eq. alpha}
\end{equation}

It should be remarked that Eq.~\eqref{eq. R pwa} assumes infinite dimensions of the porous absorber~\citep{brandao2012estimation}. Therefore, the validity of the sound absorption estimate via Eq.~\eqref{eq. alpha} does not hold for finite absorbers due to the edge-diffraction effect. Additionally, decreasing the distance between the sound source and the microphones introduces deviations in the lower frequencies because wave reflection cannot be assumed specular anymore~\citep{brandao2015review}.

\subsection{Boundary element method}
\label{Sec. 2d}
The BEM in this work follows the experimental setup shown in Fig.~\ref{fig: figure_01} and the considerations that were made by Zea et al.~\citep{zea2023sound} and Brand\~{a}o et al.~\citep{brandao2012estimation}. With this model, it is possible to simulate the impedance above the surface of a finite and locally reacting porous material~\citep{brandao2012estimation}. Furthermore, the BEM model allows it to model the edge diffraction effect caused by the finiteness of the sample of interest~\citep{hirosawa2009comparison}. 
A rectangular sample is flush-mounted to an infinite rigid baffle, as seen in Fig.~\ref{fig: figure_01}. The surface impedance $Z_{s}$ of the sample builds the boundary condition for the sample. The sample is meshed on its top surface and not along its thickness. Using the material model by Miki~\citep{miki1990acoustical}, the characteristic impedance $Z_{c}$ and the propagation constant $k_{p}$ of a porous material can be estimated by 
\begin{equation}
    Z_{c}(f) = \rho_{0}c_{0}(1+5.50\zeta^{-0.632} - \ima 8.43\zeta^{-0.632}),
    \label{eq. miki Zc}
\end{equation}
\begin{equation}
    k_{p}(f) = k_{0}(1+7.81\zeta^{-0.618} - \ima 11.41\zeta^{-0.618}),
    \label{eq. miki kp}
\end{equation}
with 
\begin{equation}
    \zeta \coloneqq \zeta(f,\sigma) = \frac{f}{\sigma \cdot 10^3},
    \label{eq. miki xi}
\end{equation}
where $\sigma$ describes the flow resistivity of the material and $\rho_{0}$ the density of air. From this, the surface impedance is calculated by 
\begin{equation}
    Z_{s}(f,\theta) = -\ima \frac{Z_{c}}{\cos{(\theta_{t})}}\cot{[k_{p}d\cos(\theta_{t})]}
    \label{eq. allard zs},
\end{equation}
with $\theta_{t} = \arcsin{[\frac{k_{0}}{k_{p}}\sin{(\theta)}]}$ and $d$ as the sample thickness~\citep{allard2009propagation}. 
With this boundary condition, the sound pressure $p(\textbf{r})$ at the receiver position can be written as the Helmholtz/Huygens integral 
\begin{equation}
    c(\textbf{r})p(\textbf{r}) = \frac{\e^{-\ima k_{0}||\textbf{r}-\textbf{r}_{q}||}}{||\textbf{r}-\textbf{r}_{q}||} + \frac{\e^{-\ima k_{0}||\textbf{r}-\textbf{r}'_{q}||}}{||\textbf{r}-\textbf{r}'_{q}||} 
    - \frac{\ima k_{0}}{Z_{s}} \int_S p(\textbf{r}_{S})  
    \frac{\e^{-\ima k_{0}||\textbf{r}-\textbf{r}_{S}||}}{4\pi||\textbf{r}-\textbf{r}_{S}||} \,\text{d}S,
    \label{eq. BEM cont}
\end{equation}
where $\textbf{r}_{S}$ can be any point on the sample surface, $\textbf{r}'_{q}$ is again the image sound source, and $S$ denotes the surface boundary of the absorbing sample~\citep{brandao2012estimation}. If the receiver is placed on the sample surface $c(\textbf{r})$ is 0.5, otherwise it is 1.0. Discretizing the sample surface into $N$ elements and applying a boundary element mesh with piecewise constant sound pressure leads to 
\begin{equation}
    c(\textbf{r})p(\textbf{r}) = \frac{\e^{-\ima k_{0}||\textbf{r}-\textbf{r}_{q}||}}{||\textbf{r}-\textbf{r}_{q}||} + \frac{\e^{-\ima k_{0}||\textbf{r}-\textbf{r}'_{q}||}}{||\textbf{r}-\textbf{r}'_{q}||} 
    - \frac{\ima k_{0}}{Z_{s}} \sum_{n=1}^{N} p(\textbf{r}_{S_{n}})\ 
    \int_{S_{n}} \frac{\e^{-\ima k_{0}||\textbf{r}-\textbf{r}_{S_{n}}||}}{4\pi||\textbf{r}-\textbf{r}_{S_{n}}||} \,\text{d}S_{n}.
    \label{eq. BEM discrete}
\end{equation}
Placing the receiver on the sample surface for each grid position $\textbf{r} = \textbf{r}_{S_{n}}$ with $n$ being varied between 1 and $N$, the sound pressures of each element can be determined from a system of equations---the integral in Eq.~\eqref{eq. BEM discrete} is calculated with Gauss--Legendre quadrature using $36$ points per element~\cite{Wu2000, Atalla2015}. If the sound pressures of the individual elements are then reinserted into Eq.~\eqref{eq. BEM discrete} the sound pressures at the receiver positions $\textbf{r}_{i} = [0,0,z_{i}]^{\mathrm{T}}$ with $i$ $\epsilon$ \{1,2\} (in accordance with the experimental setup) can be determined~\citep{brandao2012estimation}. The use of the integration procedure mentioned above has proven to be sufficiently accurate in our case. However, for a more general use of the BEM, a more sophisticated adaptive integration technique is recommended~\cite{BookChapter3_Kaltenbacher}.

\section{Methodology}
\label{Sec. 3}

\subsection{Numerical dataset}
\label{Sec. 3a}
The generation of numerical data in this work follows the procedure proposed by Zea~et~al.~\citep{zea2023sound} and the setup shown in Fig.~\ref{fig: figure_01}. The BEM model is well-studied~\citep{brandao2012estimation} and is openly accessible at \url{https://github.com/eric-brandao/finite_bem_simulator}. The simulations were conducted with \textit{GoogleColab}. 

The two microphones are positioned above the center of the sample at $x=y=0$ and a height of $z_1 = \SI{1}{\centi\meter}$ and $z_2 = \SI{3}{\centi\meter}$, respectively. 
The frequency range considered is between \SI{100}{\hertz} and \SI{2000}{\hertz}, and the frequency resolution is \SI{10}{\hertz}, resulting in $190$ discretized frequency steps. The BEM parameters are varied according to Table~\ref{Tab. BEM}. Following Zea et al.~\citep{zea2023sound}, the thickness and flow resistivity values are drawn from log-uniform distributions to ensure an appropriate sampling of the sound absorption coefficient function. In contrast, the remaining parameters are drawn from uniform distributions. 
The parameter space is defined according to typical values for porous absorbers and covers various absorber samples and numerical setups concerning the source position. 

\begin{table}[!h]
\begin{center}
\begin{tabular}{c c c c}
 \hline
 Parameter & Unit & Value & Sampling\\
 \hline
 \hline
 Frequency $\mathbf{f}$ & [\si{Hz}] & [100:10:2000] & Uniform\\
 Sample side $L_{x}$ & [\si{cm}] & [20, 100] & Uniform\\
 Sample side $L_{y}$ & [\si{cm}] & [20, 100] & Uniform\\
 Sample thickness $d$ & [\si{mm}] & [5, 200] & Log-uniform\\
 Flow resistivity $\sigma$ & [\si{kNs/m^{4}}] & [5, 100] & Log-uniform\\
 Source distance $||\textbf{r}_{q}||$ & [\si{m}] & [1.2, 1.8] & Uniform\\
 Source azimuth $\phi$ & [\si{deg}] & [0, 360] & Uniform\\
 Source elevation $\theta$ & [\si{deg}] & [0, 80] & Uniform\\
\end{tabular}
\caption{Parameters of the BEM model which were used to generate the numerical datasets.}
\label{Tab. BEM}
\end{center}
\end{table}

A total of $50000$ samples were generated for training and validation purposes. An additional $3000$ samples were generated to test the network on unseen data. The BEM simulation can be divided into two main steps: (1) assembling the BEM matrix for a given sample geometry and (2) computing the pressure field for the given geometry in combination with the source location and material properties. In this way, it is possible to reduce the computational effort~\citep{zea2023sound}. For step (1), $530$ base cases ($500$ for training and validation and $30$ for testing) were generated by assembling the BEM matrices for sample sizes drawn randomly from the distributions shown in Table~\ref{Tab. BEM} ($L_{x}, L_{y}$). These base cases were then used in step (2) to generate the pressure fields for $100$ random and unique combinations of the remaining BEM parameters. 

Linear interpolation with Gauss-Legendre quadrature with $36$ points per element is applied to calculate the integrals in Eq.~\eqref{eq. BEM discrete}~\citep{Atalla2015,wu2002boundary}. Following the procedure described in Sec.~\ref{Sec. 2d} and using Eq.~\eqref{eq. transfer function}, the transfer function between the two microphones is computed for each sample. The corresponding sound absorption coefficient without the edge-diffraction effect is then calculated using Eqs.~\eqref{eq. miki Zc},~\eqref{eq. miki kp},~\eqref{eq. allard zs},~\eqref{eq. alpha} and the relation $R = (Z_{s}\cos(\theta)-\rho_{0} c_{0})/(Z_{s}\cos(\theta)+\rho_{0} c_{0})$. The reference absorption coefficient with edge diffraction given by the traditional two-microphone method is calculated with Eqs.~\eqref{eq. R pwa} and~\eqref{eq. alpha}. \\

\subsection{Experimental dataset}

Free-field measurements of the absorption coefficient of two samples of glass wool with flow resistivity $54.7 \pm 8.8$~\si{kN s m^{-4}} (Focus A 2 cm, Saint-Gobain Ecophon, Hyllinge, Sweden) were performed in an anechoic chamber at the Marcus Wallenberg for Sound and Vibration Research (KTH) using the two-microphone method described in Sec.~\ref{Sec. 2b}. Two sample sizes were considered: $600\times600\times~20$~\si{mm^3} and $300\times600\times20$~\si{mm^3}. For the latter, a regular sample of $600\times600\times20$~\si{mm^3} was cut in half using a hand saw. 

\begin{figure}[!h]
\includegraphics[width=0.4\textwidth]{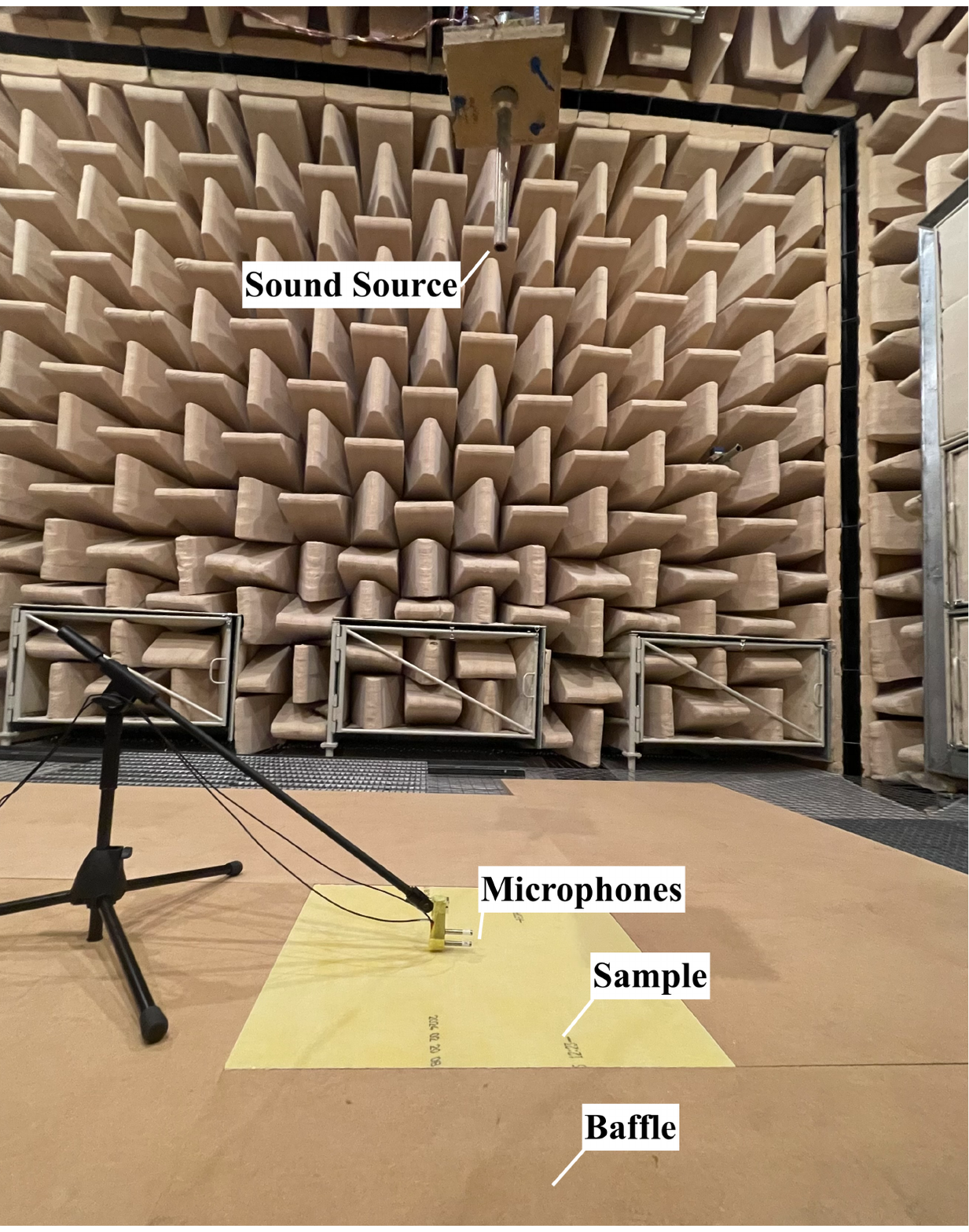}
\centering
\caption{Experimental setup of the two-microphone method in the anechoic chamber at the Marcus Wallenberg Laboratory for Sound and Vibration in Stockholm. }
\label{'fig: figure_02'}
\end{figure}

The sample was placed above a \SI{10}{\milli\meter}-thick plywood rigid backing, together forming a \SI{3}{\centi\meter}-thick layer that was then enclosed within a rigid baffle made with four $2 \times 1 \times 0.03$~m$^3$ medium-density fiberboard (MDF) panels. The sample was flush-mounted, and care was taken to minimize slits between the sample and the baffle, as well as slits between the MDF panels themselves. 

Two {G.R.A.S. 40PH} microphones were positioned above the center of the sample at heights of $z=\SI{1}{\centi\meter}$ and $z=\SI{3}{\centi\meter}$, respectively. A {Monacor KU-516} monopole sound source played an exponential sine sweep signal of \SI{10}{\second} duration with a \SI{12.8}{\kilo\hertz} sampling frequency using a {Brüel~\&~Kjær Type 2706} amplifier and a {National Instruments NI eDAQ-9178} digital acquisition system. The cut-on and cut-off frequencies of the sweep were set to \SI{50}{\hertz} and \SI{4050}{\hertz}, respectively. The impulse responses at each microphone were computed, and then, the transfer function $H_{c}$ between the two microphones (with the one at \SI{1}{\centi\meter} above a rigid surface as reference). The source position $\|\mathbf{r}_q\|$ and elevation angle $\theta$ were varied with respect to the sample center. The source azimuth was held constant at $0^\circ$. The source position was measured with a double meter stick in relation to the sample center, and $\theta$ was calculated by trigonometry. Table~\ref{Tab. Measurements} gives a summary of the investigated experimental setups.

\begin{table}[!h]
\begin{center}
\begin{tabular}{c c c c c}
 Measurement & $L_{x}$ [m] & $L_{y}$ [m] & $\left\Vert \mathbf{r}_q \right\Vert$ [m] & $\theta$ [deg] \\ 
 \hline
 \hline
\Romannum{1} & 0.6 & 0.6 & 1.21 & 0.0 \\
\Romannum{2} & 0.6 & 0.6 & 1.46 & 0.0 \\
\Romannum{3} & 0.6 & 0.6 & 1.64 & 0.0 \\
\Romannum{4} & 0.6 & 0.6 & 1.70 & 30.0 \\
\Romannum{5} & 0.3 & 0.6 & 1.20 & 0.0 \\
\Romannum{6} & 0.3 & 0.6 & 1.51 & 0.0 \\
\Romannum{7} & 0.3 & 0.6 & 1.66 & 0.0 \\
\Romannum{8} & 0.3 & 0.6 & 1.69 & 27.0 \\

\end{tabular}
\caption{Measurements and parameters considered for the experimental datasets. }
\label{Tab. Measurements}
\end{center}
\end{table}

The measured transfer functions were post-processed by adjusting the frequency range and resolution. Two subsequent moving-average filters, with each a window size of $20$ frequency steps, were applied to smooth out remaining artifacts caused by slits in the baffle and measurement noise in the transfer functions. Equation~\eqref{eq. R pwa} and~\eqref{eq. alpha} were used to calculate the sound absorption coefficient for the traditional two-microphone method. 

\subsection{Proposed neural network}
\label{Sec. 3c}
The here presented deep neural network predicts the angle-dependent sound absorption coefficient \textit{in situ} for an infinite sample size based on the measurement of a finite sample. The network estimates the absorption coefficients using the measured transfer function between the microphones and the wave incidence angle as inputs. Unlike the network proposed by Zea et al.~\citep{zea2023sound}, which has used the absolute sound pressure values as input, the proposed network considers the complex-valued transfer function of the recorded sound pressures. In this way, similarly to M\"ueller et al.~\citep{muellergiebeler2024}, the network exploits the phase content of the measurements. 

\subsubsection{Architecture}
\begin{figure*}[!ht]
    \includegraphics[width=0.9\textwidth]{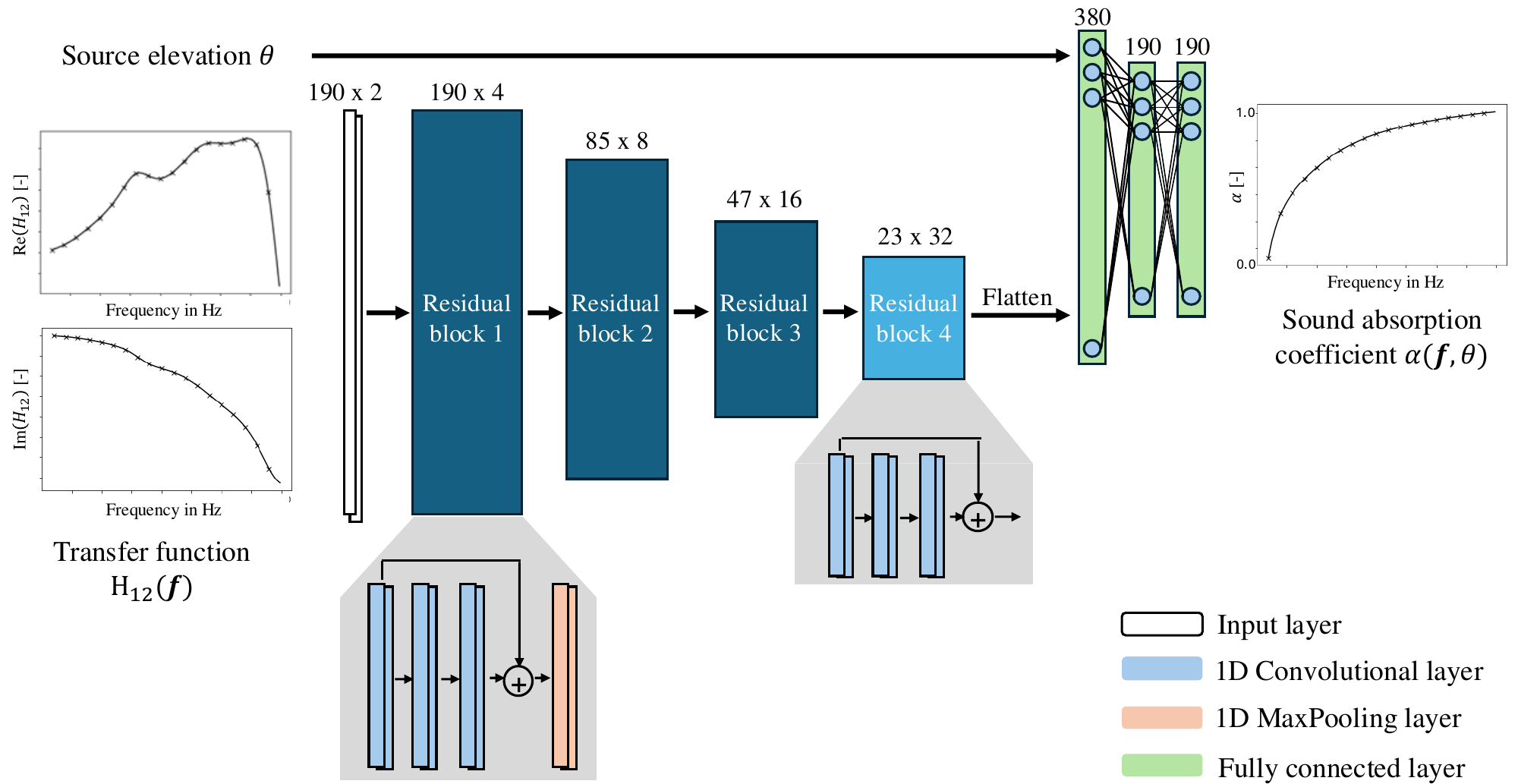}
    \centering
    \caption{Schematic of the network architecture. The total number of trainable parameters is 406300. }
    \label{fig: figure_03}
\end{figure*}

The network architecture, shown in Fig.~\ref{fig: figure_03}, can be divided into a fully convolutional encoder followed by a fully-connected part for decoding. The network receives two inputs: (1) the complex-valued transfer function $H_{12}(\mathbf{f})$ and (2) the source elevation $\theta$. The network’s output is the absorption coefficient $\alpha(\textbf{f})$. 

Following existing research on frequency-dependent and complex-valued input vectors~\citep{Eser2023,eser4452034hybrid}, a 1D-convolutional architecture was chosen to process the transfer function input. The real and imaginary parts of the transfer function $H_{12}(\textbf{f})$ are treated as separate input feature channels. The discrete input is given by $[\Re\{H_{12}(\textbf{f})\}, \Im\{H_{12}(\textbf{f})\}] \in \mathbb{R}^{2\times 190}$. The advantages of convolutional layers justify the choice of a fully convolutional encoder. Convolutional layers allow learning features on different detail resolutions from the sequence~\citep{geron2022hands}. Additionally, the feature extraction in convolutional layers happens across all feature channels simultaneously~\citep{dumoulin2016guide}. In this way, the network learns the real and imaginary parts of the transfer function separately and their inter-dependency, resulting in an effective and parameter-saving way of learning the complex features. 

The encoder is structured into four consecutive residual blocks, doubling the number of feature channels per block. The first three residual blocks consist of three convolutional layers and a max-pooling operation. Each residual block is implemented by adding a skip connection from the first convolutional layer to the max-pooling layer. This kind of strategy has been shown to improve learning compared to traditional convolutional neural networks~\cite {zea2023sound}. All convolutional layers have a stride of $1$ and a kernel size of $5$, and zero padding is used to maintain the sequence length within the residual block. The subsequent max-pooling operation halves the sequence length before the sequence is passed to the next residual block. This reduction in sequence length and concurrent increase in feature channels is a widely used approach in deep learning architectures~\citep{he2016deep,ronneberger2015u,simonyan2014very}. 

The fourth and last residual block is similar to the first three blocks but does not apply the max-pooling operation at the end. 
It builds the transition from the convolutional part to the fully connected part of the architecture and constitutes the bottleneck. The output of the last residual block is flattened and concatenated with the scalar input of the elevation angle, forming the input to the subsequent three fully connected layers. The first fully connected layer has $380$ neurons, and the last two layers have $190$ neurons to match the output sequence length to the input sequence length. The final fully-connected layer produces the output of the network. A sigmoid activation function is chosen to incorporate the physical knowledge about the absorption coefficient, which ranges from zero to one. The remaining fully connected layers and convolutional layers used a tanh activation function. Glorot uniform weight initialization is adopted to minimize vanishing gradients~\citep{glorot2010}. 

Various network architecture configurations were compared in the network optimization process, and the network’s generalization abilities were evaluated. The generalization performance was evaluated over the parameter space of the numerical simulations, consisting of six hyperparameters: (1) the sample size, (2) the sample thickness, (3) the sample’s flow resistivity, (4) the source distance, (5) elevation, and (6) azimuth. The hyperparameter analysis showed that adding the elevation angle $\theta$ in the latent space of the neural network improved the prediction performance for high wave incidence angles. The overall training and validation loss decreased by approximately one order of magnitude. 

\subsubsection{Training setup and loss function}
The neural network training was performed using the numerical dataset of $50000$ samples (see Sec.~\ref{Sec. 3a}). Before training, the dataset was randomly split into a training set and a validation set with a ratio of 80:20. While the training set is used to optimize the network parameters, the validation set’s purpose is to monitor the network's performance with unseen data. The inputs were standardized to improve the convergence rate and optimization stability~\citep{geron2022hands}. For this purpose, the mean $\mu$ and standard deviation $s$ per feature were calculated over all samples in the training set. In this way the standardized inputs for the transfer function $H_{12}$ and the incidence angle $\theta$ follow as 
\begin{equation}
    \Re{(H_{12_{n}}(f_{m}))}' = \frac{\Re{(H_{12_{n}}(f_{m}))} - \mu_{\Re,m}}{s_{\Re,m}},
    \label{eq. normalization H}
\end{equation}

\begin{equation}
    \Im{(H_{12_{n}}(f_{m}))}' = \frac{\Im{(H_{12_{n}}(f_{m}))} - \mu_{\Im,m}}{s_{\Im,m}},
    \label{eq. normalization H}
\end{equation}

\begin{equation}
    \theta'_{n} = \frac{\theta_{n} - \Tilde{\mu}}{\Tilde{s}},
    \label{eq. normalization theta}
\end{equation}
where $n$ is the index of the sample and $m$ is the index of the discrete frequency. The samples in the validation and test sets were standardized, using the mean and the standard deviation computed from the training set. 

The network was trained for 250 epochs and employed on the validation set after each epoch. The mean squared error (MSE) was used as the performance metric. The optimization was performed in mini-batches of size $64$ and using an Adam optimizer with a weight decay of $\lambda = 10^{-3}$, as Scikit-Learn recommends~\citep{geron2022hands}. Therefore, the complete loss term follows as 
\begin{equation}
    \text{MSE} = \frac{1}{N} \frac{1}{190} \sum_{n=1}^N \sum_{m=1}^{190} (\alpha_{n}(f_{m}) - \alpha_{n}(f_{m}))^2 + \lambda \sum_{k=1}^{K}w_{k}^2,
    \label{eq. loss}
\end{equation}
where $N$ is the number of samples and $K$ the number of parameters $w$. 
The initial learning rate was set to $10^{-3}$ and, from the eleventh epoch on, was reduced exponentially by $0.9$. Early stopping was applied to prevent overfitting of the network. The network was implemented and trained using Python 3.12 and Tensorflow-Keras 3.3.3.

\section{Results}
\label{Sec. 4}

\subsection{Training performance}

The training of the neural network was conducted according to Sec.~\ref{Sec. 3c} and was executed on a local machine with 16 GB RAM, an Intel\textsuperscript{\textregistered}Core\texttrademark i7-6500U CPU, and no GPU. During training, the MSE was calculated after each mini-batch of the training set (training loss) and after each epoch on the validation set (validation loss). One epoch took on average \SI{12}{\second} of computation time. The early stopping mechanism stopped the parameter optimization after 125 epochs because no further significant improvement in the validation loss was detected. 

\begin{figure}[!h]
\includegraphics[width=0.45\textwidth]{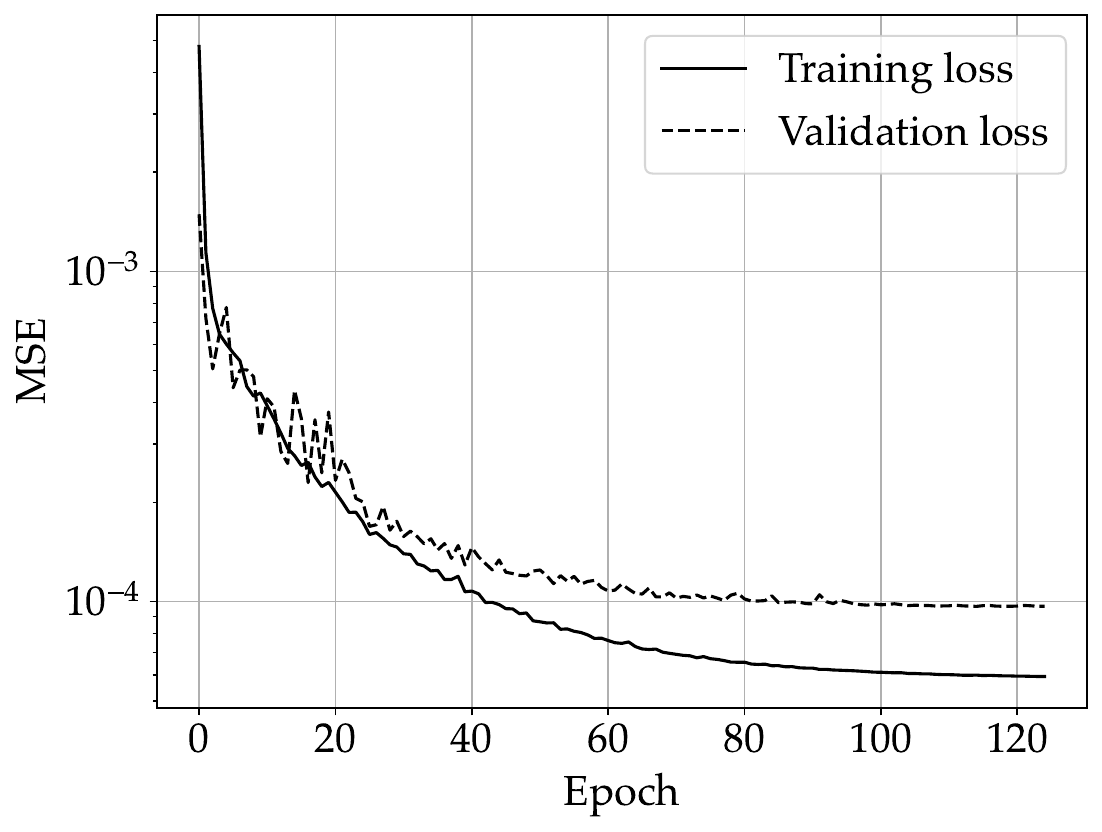}
\centering
\caption{Training and validation loss over $125$ training epochs.}
\label{fig: figure_04}
\end{figure}

Figure~\ref{fig: figure_04} shows the training and validation loss evolution over the training epochs. The mean training loss versus epoch is displayed in a logarithmic y-scale for clearer visualization. As can be seen, training and validation loss both show a very good convergence behavior with a training loss of \mbox{$5.93\cdot10^{-5}$} after $125$ epochs. The validation loss converges to a value of \mbox{$9.68\cdot10^{-5}$}, indicating that the network generalizes well to unseen data. Based on the validation loss curve, there is no indication of overfitting. 

\subsection{Numerical validation}

\begin{figure}[!h]
    \includegraphics[width=0.45\textwidth]{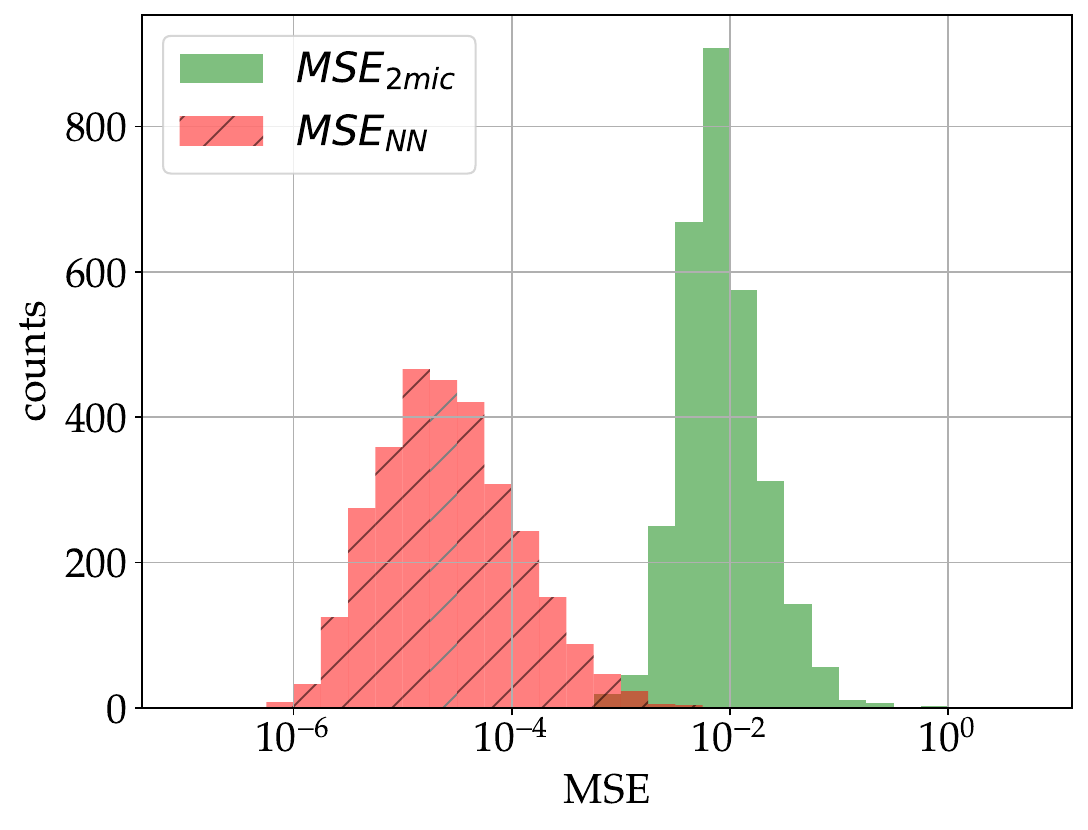}
    \centering
    \caption{Histograms of the recorded error per sample for the two-microphone method and for the neural network predictions using the numerical test set. }
    \label{fig: figure_05}
\end{figure}

The performance of the proposed method is first evaluated using the numerical test set, which comprises $3000$ unique BEM simulations of finite porous absorbers (see Sec~\ref{Sec. 3a}). The MSE over all test samples is \mbox{$8.42~\cdot~10^{-5}$}.  
Figure~\ref{fig: figure_05} summarizes the results in a histogram as a comparison between the errors per sample for the traditional two-microphone method ($\text{MSE}_{2mic}$) and the neural network predictions ($\text{MSE}_{NN}$). The reference for the classical two-microphone and proposed methods is obtained by a transfer matrix approach using the Miki material model of a porous absorber on a rigid backing. 

The error distribution for the neural network shows a clear shift to lower values. The mean error over all samples for the neural network is about three orders of magnitudes lower than the one for the traditional two-microphone method. Furthermore, the highest observed error of the neural network for the numerical test set is in the range of $10^{-2}$, which is still lower than the average error of the traditional two-microphone method. This superior performance indicates that the proposed network effectively mitigates the deviations in the sound absorption coefficient introduced by edge diffraction. Furthermore, the results for the numerical test set illustrate the generalization abilities of the neural network for different material characteristics, sample sizes, and experimental setups (see the considered parameter spaces in Table~\ref{Tab. BEM}). It should be emphasized that the network shows a stable prediction performance for sample sizes down to \SI{30}{\centi\meter} by \SI{30}{\centi\meter} and source elevation angles up to $80^\circ$. However, the edge-diffraction effect is especially prominent for those samples. Only for sample sizes with one edge being shorter than \SI{30}{\centi\meter}, a slight increase in the MSE of about \mbox{$4~\cdot~10^{-4}$} was observed. 

\subsection{Experimental validation}

In this section, the performance of the network are evaluated experimentally based on free-field measurements using the two-microphone method described in Sec.~\ref{Sec. 2b}. 

\subsubsection{Validation with BEM predictions}
\begin{figure}[!h]
    \centering
    \begin{subfigure}{0.23\textwidth} 
        \includegraphics[width=\linewidth]{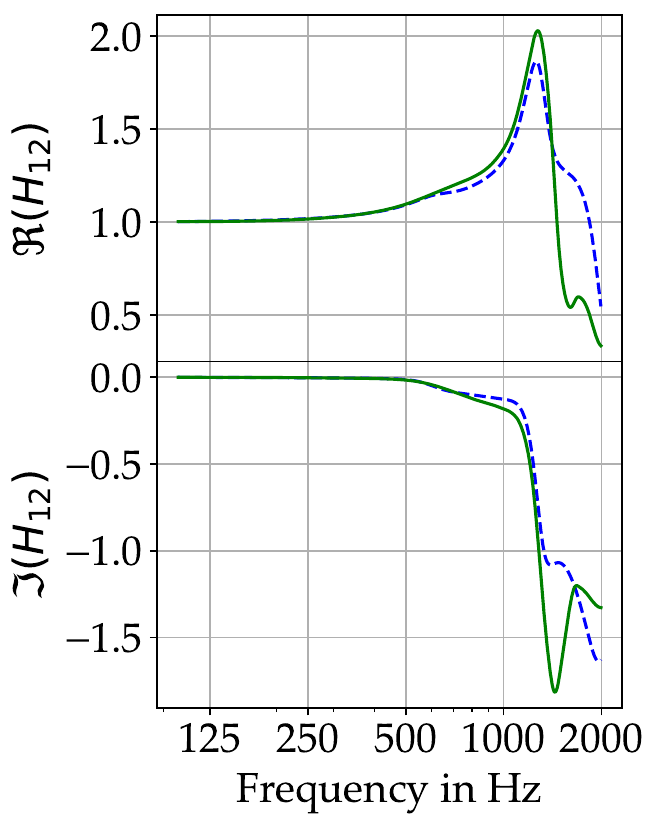} 
        \caption{}
        \label{fig:subfig6a}
    \end{subfigure}
    \hfill 
    \begin{subfigure}{0.23\textwidth}
        \includegraphics[width=\linewidth]{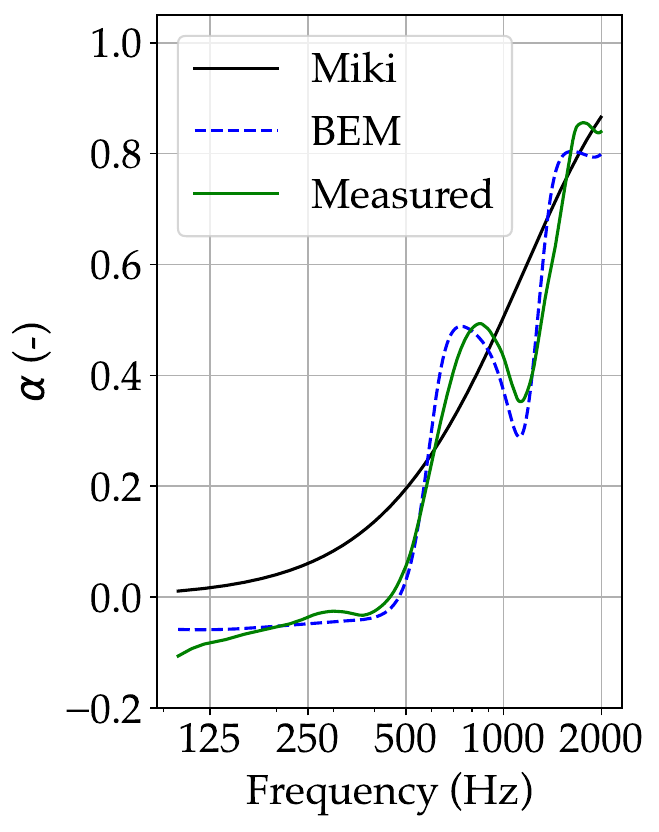} 
        \caption{}
        \label{fig:subfig6b}
    \end{subfigure}
    \begin{subfigure}{0.23\textwidth} 
        \includegraphics[width=\linewidth]{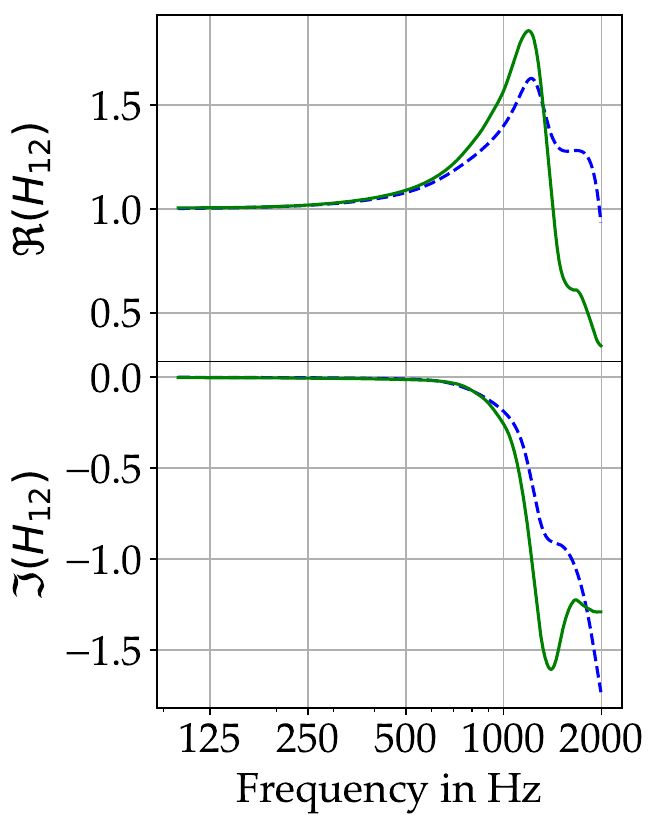} 
        \caption{}
        \label{fig:subfig6c}
    \end{subfigure}
    \hfill 
    \begin{subfigure}{0.23\textwidth}
        \includegraphics[width=\linewidth]{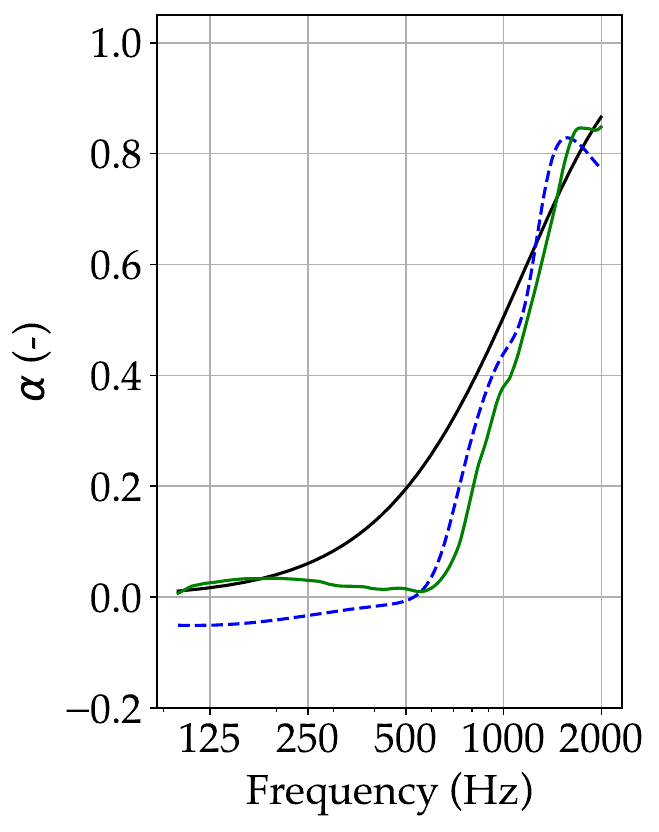} 
        \caption{}
        \label{fig:subfig6d}
    \end{subfigure}
    
    \caption{Comparison between measurement \Romannum{1} (upper row) and measurement \Romannum{5} (lower row) in green to the corresponding BEM simulations in dashed blue: The transfer functions in (a) and (c) and the calculated absorption coefficients in (b) and (d). Solid black curves in (b) and (d) show expected values for an infinite sample.}
    \label{fig:figure_06}
\end{figure}

Before evaluating the network predictions, the obtained transfer functions are validated with BEM simulations. Figure~\ref{fig:figure_06} shows the transfer functions for measurements \Romannum{1} and \Romannum{5} (see Table \ref{Tab. Measurements}), along with sound absorption curves derived using the traditional two-microphone method. BEM simulations for the given sample sizes and source positions (dashed blue), using the Miki model with the estimated flow resistivity of $54.7$~\si{kN s m^{-4}} as the boundary condition, are displayed supplementary.

The measured and simulated transfer functions align well up to \SI{1400}{\hertz}. For higher frequencies, the measured real and imaginary parts diverge, displaying a sharp peak and a steep decline, respectively (Figs.~\ref{fig:subfig6a} and~\ref{fig:subfig6c}). These deviations affect the calculated sound absorption coefficients in Figs.~\ref{fig:subfig6b}~and~\ref{fig:subfig6d}, resulting in slightly shifted absorption curves. Similar results were obtained for all measurements of the large and the small samples, respectively, and are not shown here for brevity. The measured transfer functions of the small sample under normal wave incidence with different source heights showed a higher variability between measurements. Although there are deviations between the simulated and the measured transfer functions, the BEM model is validated as a suitable simulation method. Uncertainties in the estimation of the flow resistivity, sample-specific variations, and experimental uncertainties could explain the observable differences. 

The plots of the sound absorption coefficients in Figs.~\ref{fig:subfig6b}~and~\ref{fig:subfig6d} illustrate the edge-diffraction effect. The results for measurement scenario \Romannum{1} show the characteristic oscillations around the expected sound absorption of an infinite-sized sample given by the Miki model in black. BEM simulation and measurement yield physically incorrect, negative absorption values for frequencies below \SI{500}{\hertz}. Similar results can be observed for measurement \Romannum{5}. Although the measurement results for very low frequencies align better with the Miki model, no meaningful interpretation of the absorption behavior is possible for frequencies between \SI{250}{\hertz} and \SI{1500}{\hertz}, emphasizing the need for an improved absorption inference method. 

\subsubsection{Validation at normal incidence}

This work uses two separate characterization methods to validate the proposed method.
A reference sound absorption curve is first calculated using the Delany-Bazley-Miki model~\citep{miki1990acoustical} and the flow resistivity value provided by the manufacturer. 
Secondly, impedance tube measurements have been conducted on samples from the same batch of panels used for the two-microphone free-field setup. 
Two impedance tubes were used to test $10$ and $8$ samples, respectively. Impedance tube 1 has an inner diameter of $34.86$~\si{mm} and comprises two microphones spaced by \SI{29}{\milli\meter}. The measured absorption spectra are valid in the range $f\in[300,5000]~\si{Hz}$ for this setup. Impedance tube 2 has an inner diameter of 100 mm and uses a single microphone moved between two positions 90~\si{mm} apart, leading to a valid frequency range of $f\in[100,1700]~\si{Hz}$.
The measured sound absorption spectrum is presented as a mean value and standard deviation over the different samples.

\begin{figure*}[!t]
    \centering
    \begin{subfigure}{0.32\textwidth} 
        \includegraphics[width=\linewidth]{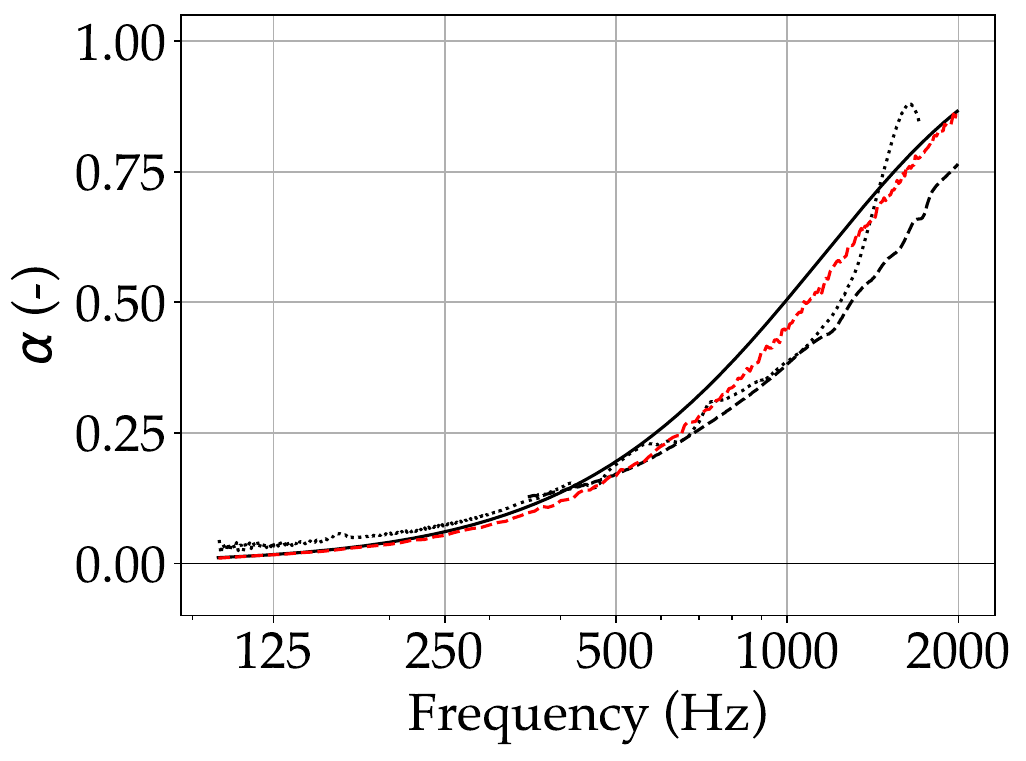} 
        \caption{}
        \label{fig:subfig7a}
    \end{subfigure}
    \hfill 
    \begin{subfigure}{0.32\textwidth}
        \includegraphics[width=\linewidth]{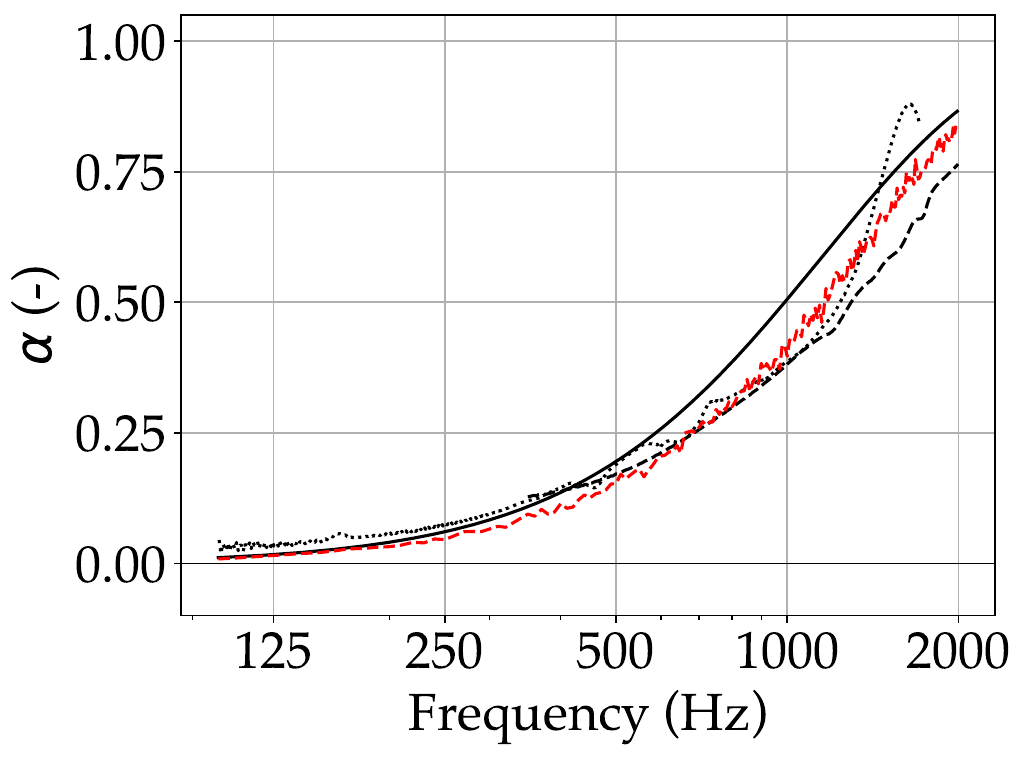} 
        \caption{}
        \label{fig:subfig7b}
    \end{subfigure}
    \hfill
    \begin{subfigure}{0.32\textwidth}
        \includegraphics[width=\linewidth]{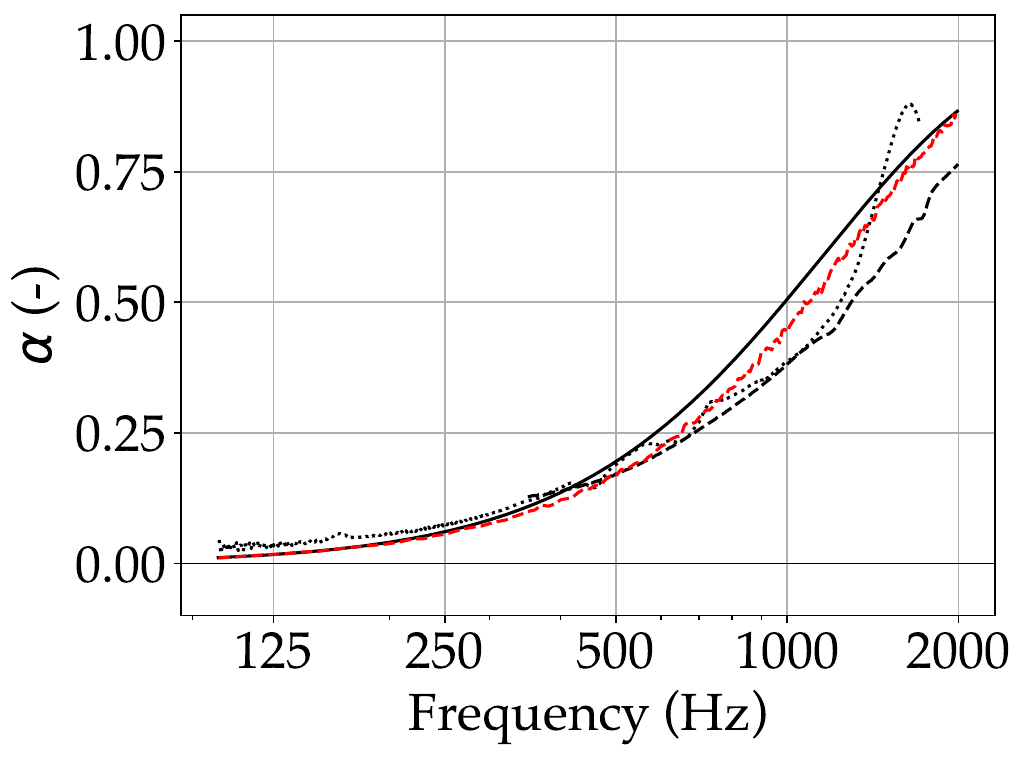} 
        \caption{}
        \label{fig:subfig7c}
    \end{subfigure}
    
    \vspace{0.5em} 
    
    \begin{subfigure}{0.32\textwidth}
        \includegraphics[width=\linewidth]{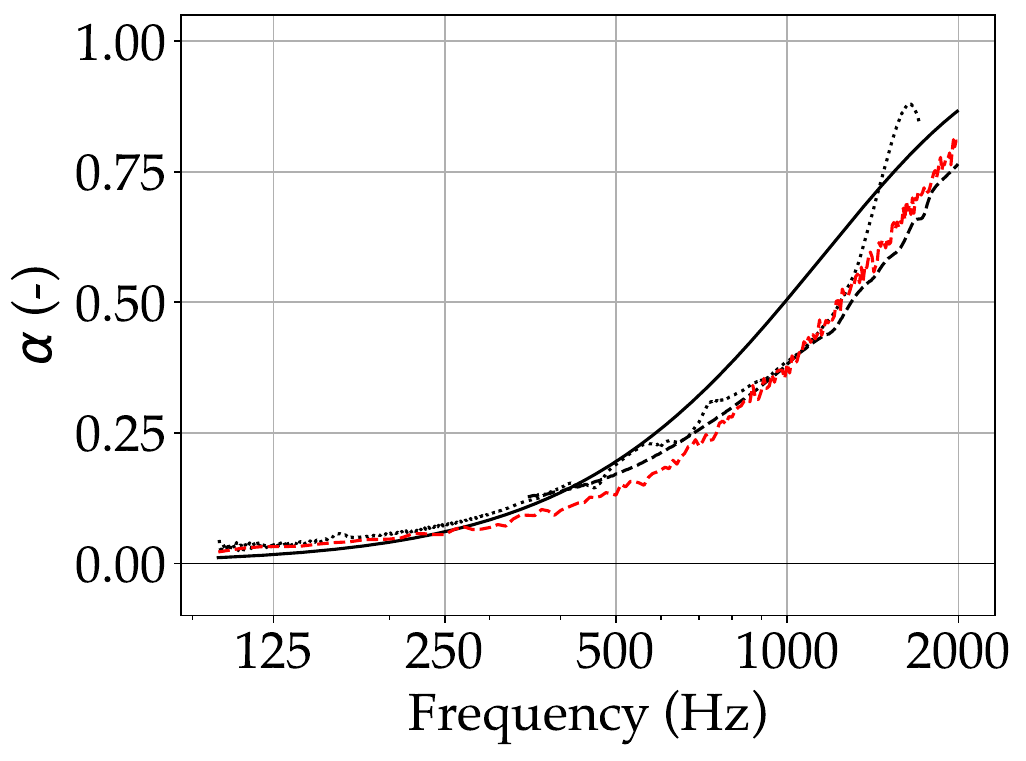} 
        \caption{}
        \label{fig:subfig7d}
    \end{subfigure}
    \hfill
    \begin{subfigure}{0.32\textwidth}
        \includegraphics[width=\linewidth]{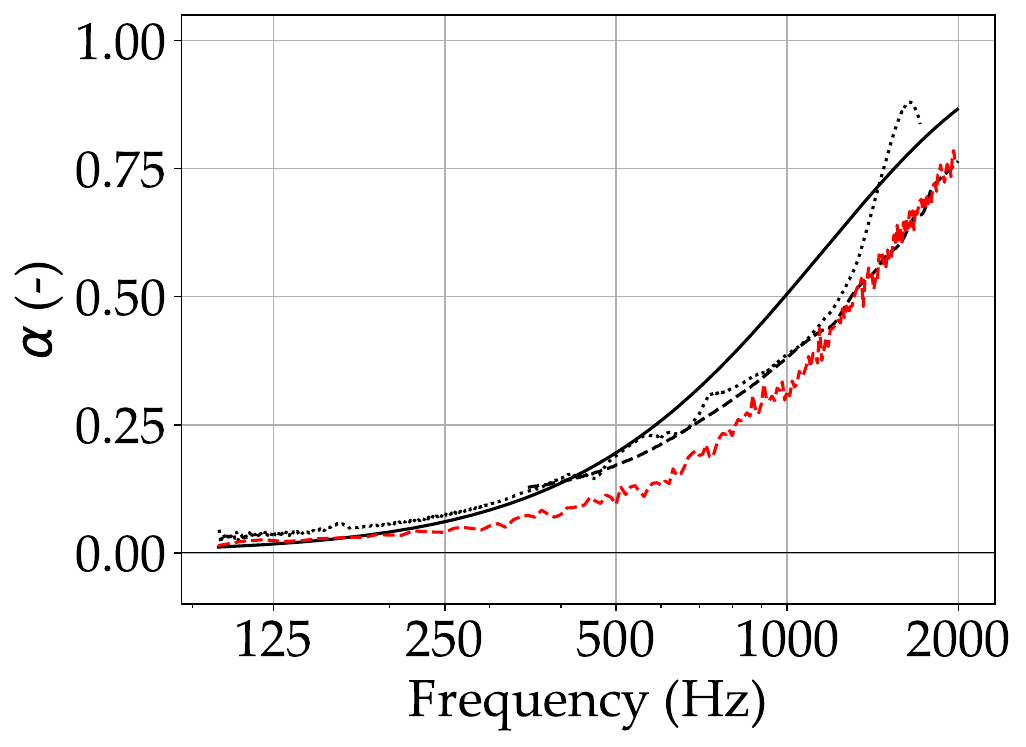} 
        \caption{}
        \label{fig:subfig7e}
    \end{subfigure}
    \hfill
    \begin{subfigure}{0.32\textwidth}
        \includegraphics[width=\linewidth]{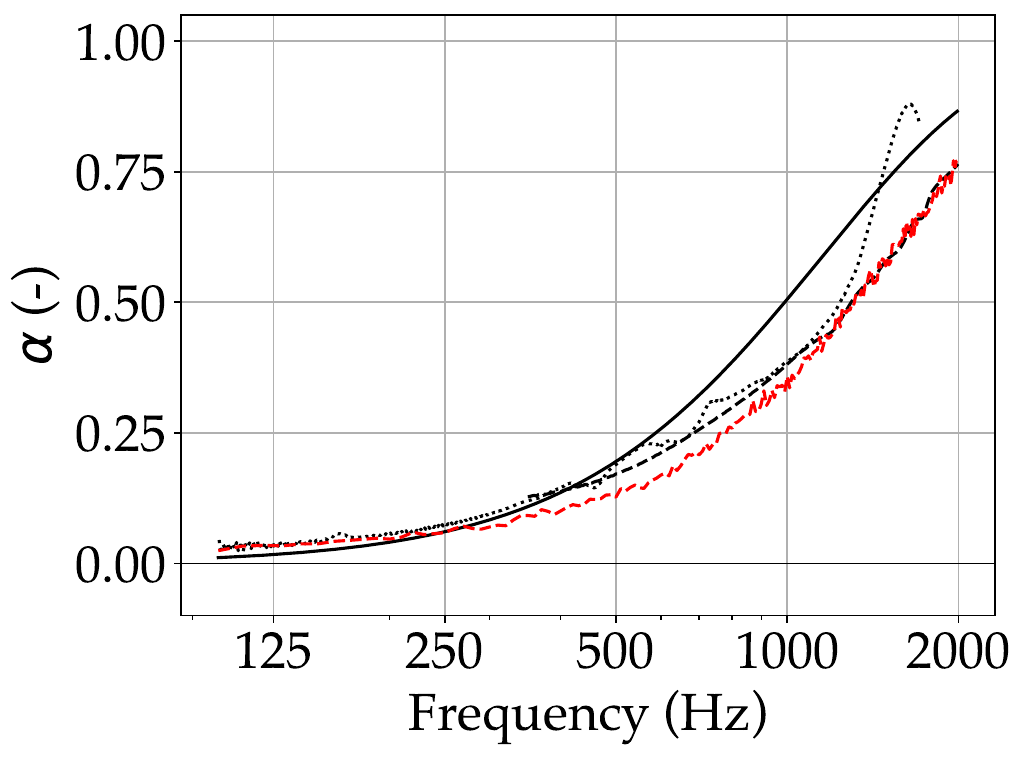} 
        \caption{}
        \label{fig:subfig7f}
    \end{subfigure}
    
    \caption{Preliminary: Comparison of the results for the normal sound absorption coefficient given by the neural network (dashed red), 
    the Miki model (black), and the impedance tube measurements (dashed black/dotted black). a) 60~\si{cm}~$\times$~60~\si{cm},~$\left\Vert \mathbf{r}_q \right\Vert=$~1.21~\si{m}, b) 60~\si{cm}~$\times$~60~\si{cm},~$\left\Vert \mathbf{r}_q \right\Vert=$~1.46~\si{m}, c) 60~\si{cm}~$\times$~60~\si{cm},~$\left\Vert \mathbf{r}_q \right\Vert=$~1.64~\si{m}, d) 30~\si{cm}~$\times$~60~\si{cm},~$\left\Vert \mathbf{r}_q \right\Vert=$~1.20~\si{m}, e) 30~\si{cm}~$\times$~60~\si{cm},~$\left\Vert \mathbf{r}_q \right\Vert=$~1.51~\si{m}, f) 30~\si{cm}~$\times$~60~\si{cm},~$\left\Vert \mathbf{r}_q \right\Vert=$~1.66~\si{m}.}
    \label{fig:figure_07}
\end{figure*}

Figure~\ref{fig:figure_07} shows the network predictions for all measurements of the large and the small samples with normal wave incidence, compared to the estimated sound absorption by the Miki model and the measured sound absorptions in the impedance tubes. The reference measurements with the impedance tube in this study showed a higher variability across different panels of the same material (averaged standard deviation of $13$\%) compared to measurements of the same panel (averaged standard deviation of $8$\%). These sample-specific variations in the sound absorption are also visible in the differences between the dashed black (average of one measured panel) and dotted black curves (average across multiple measured panels). Together with the knowledge about the standard deviation of about $16$\% in the acquisition of the flow resistivity, the expected sound absorption of the material under observation is likely to vary between the references given by the Miki model (in solid black) and the two impedance tube setups (in dotted and dashed black, respectively). 

The network predictions for the large sample in Figs.~\ref{fig:subfig7a}-\ref{fig:subfig7c} are consistent for the different source distances with only minimal deviations between the predicted sound absorption coefficients. For all three measurements, the network predicts a sound absorption coefficient slightly below that produced by the Miki model. The predictions show almost no fluctuations. Compared to the sound absorption coefficient given by the traditional two-microphone method in Fig.~\ref{fig:subfig6b}, the neural network compensates well for the edge-diffraction effects inherent in the measured transfer functions. The average standard deviation across multiple measurements of $3.5$\% is significantly lower than for the impedance tube measurements. 

The predictions for the small sample are illustrated in Fig.~\ref{fig:subfig7d}-\ref{fig:subfig7f}. On average, the predicted sound absorption for the small sample is lower compared to the large sample and more closely fits the sound absorption obtained via impedance tube measurements. However, the overall deviation between the predictions of the large and small samples is still in the range that can be expected based on the results of the impedance tube measurements, the nominal material characterization, and the correspondingly observed standard deviations. 
Nonetheless, the predictions for the small sample show a higher variability for the different source heights than for the large sample. Measurement \Romannum{5} (Fig.~\ref{fig:subfig7d}) and \Romannum{7} (Fig.~\ref{fig:subfig7f}) yield very similar predicted sound absorptions. The prediction for measurement \Romannum{6} (Fig.~\ref{fig:subfig7e}) deviates especially for mid-range frequencies compared to the other two predictions. 

Although the acquired transfer functions varied more strongly for different source heights compared to the measurements of the large sample, a careful investigation of the result in Fig.~\ref{fig:subfig7e} was needed. It was found that deviations of the transfer functions in the low-frequency range translate to an overall deviation in the network prediction. This effect is amplified by the real parts of the measured transfer function converging towards zero. Due to the input standardization, a slight variation in very low frequencies can lead to significantly different inputs. The network's training was solely performed on numerical data for which measurement noise and possible uncertainties were not considered. For this reason, measurement uncertainties pose a generalization challenge to the network. Future research should focus on enhancing the training process by including noisy training data. Furthermore, an enhanced input standardization could reduce the low-frequency artifacts in the predictions. Nevertheless, the averaged standard deviation across measurements \Romannum{5}-\Romannum{7} of $6.7$\% is still competitive with the standard deviation observed in the impedance tube measurements. 

\subsubsection{Validation at oblique incidence}
\begin{figure}[!h]
    \centering
    \begin{subfigure}{0.45\textwidth} 
        \includegraphics[width=\linewidth]{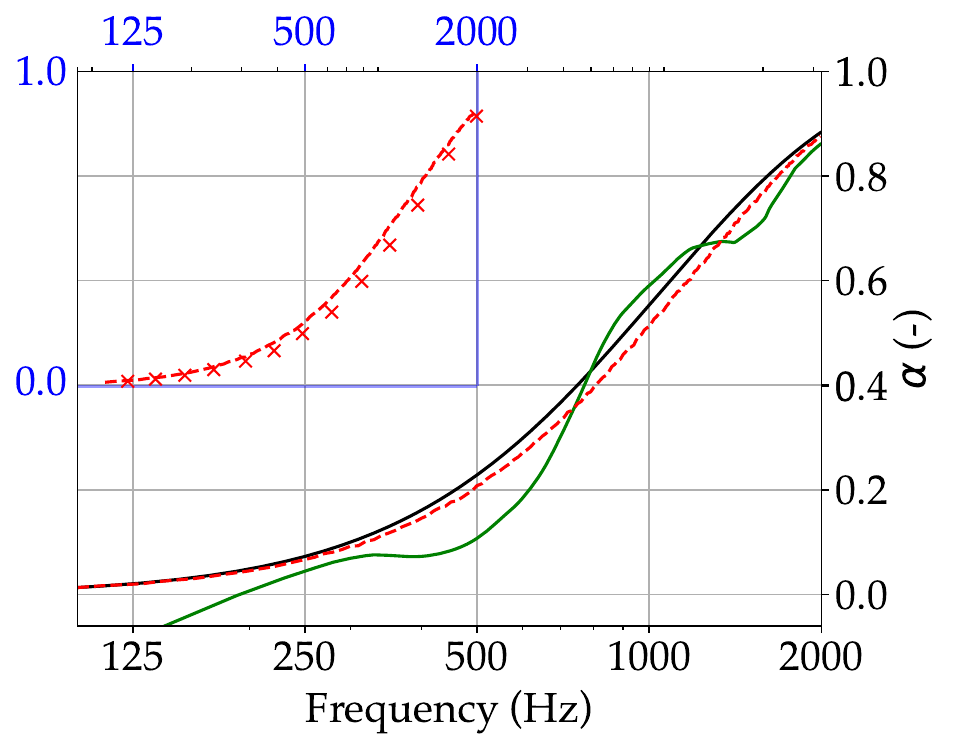} 
        \caption{}
        \label{fig:figure_08a}
    \end{subfigure}
    \hfill 
    \begin{subfigure}{0.45\textwidth}
        \includegraphics[width=\linewidth]{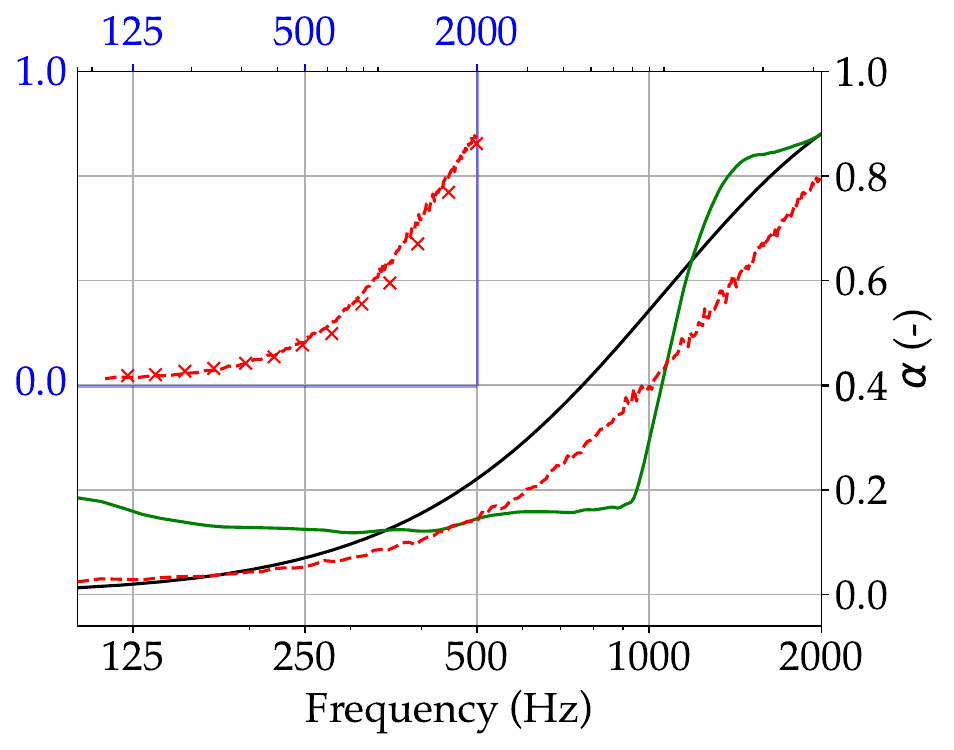} 
        \caption{}
        \label{fig:figure_08b}
    \end{subfigure}
    
    \caption{Comparison of the network predictions of measurements \Romannum{4} (a) and \Romannum{8} (b) in dashed red to the corresponding sound absorption coefficients given by the DBM model (black) and the traditional two-microphone method (green). Additional comparison to normal incidence absorption in the upper-left corner with predictions for measurements \Romannum{3} (a) and \Romannum{7} (b) as reference (red crosses). }
    \label{fig:figure_08}
\end{figure}

Lastly, the network performance is analyzed for the two oblique measurements \Romannum{4} and \Romannum{8}. Figure~\ref{fig:figure_08} compares the sound absorption coefficients estimated by the traditional two-microphone method, the network predictions, and the estimation of the Miki model for source elevations of $30^{\circ}$ and $27^{\circ}$, respectively. The sound absorption coefficient calculated with the traditional two-microphone method yields less strong oscillations for the large sample (Fig.~\ref{fig:figure_08a}) compared to the normal wave incidence scenario before, allowing a more accurate investigation of the sound absorption. The network prediction in dashed red aligns almost perfectly between the peaks and valleys of the green curve while again slightly underestimating the sound absorption given by the Miki model for the material characterization. The result agrees with the network predictions for the large sample and normal wave incidence. As visualized in the upper left corner of Fig.~\ref{fig:figure_08a}, the predicted sound absorption for an incidence angle $\theta = 30^{\circ}$ (dashed red) shows a shift towards lower frequencies compared to the prediction for normal wave incidence (red crosses). This shift is of a similar magnitude as the Miki model predicts it. 

The predicted sound absorption coefficient for the small sample and a source elevation of $27^{\circ}$ (Fig.~\ref{fig:figure_08a}) is lower across the whole frequency spectrum than the large sample’s prediction (Fig.~\ref{fig:figure_08b}).
Considering the results from Fig.~\ref{fig:figure_07}, this seems reasonable. The sound absorption was predicted to be lower than the Miki reference and in closer accordance with the impedance tube measurements for the small sample. Therefore, the results shown in Fig.~\ref{fig:figure_08b} underscore the resilience of the proposed deep-learning method against changes in the experimental setup. Compared to the normal-incidence measurement (measurement \Romannum{7}) in the upper left corner of Fig.~\ref{fig:figure_08b}, a shift in the absorption spectrum is observable that is very similar to the one observed for the large sample and the Miki model estimation. 

It should be emphasized that the input transfer functions of the two grazing-incidence measurements strongly deviate in shape from the corresponding normal-incidence measurements. For this reason, the prediction of sound absorption in close accordance with theory is not trivial, illustrating the generalization capabilities of the proposed neural network to varying source elevation. Nevertheless, the network's sensitivity towards variations of the input transfer functions in the low-frequency range described above also affects the results' interpretability. Considering the minor shift in the predicted absorption of the angled measurement of the small sample compared to the normal-incidence scenario (upper left corner of Fig.~\ref{fig:figure_08b}), the network accuracy limits the level of detail with which the angle-dependent absorption can be analyzed. Future tests on further measurements with an angled sound source will be necessary to analyze whether the predictions for varying source elevations yield similarly accurate results as for the numerical test data. 

Overall, the results in Fig.~\ref{fig:figure_07} and Fig.~\ref{fig:figure_08} validate the proposed deep-learning method with experimental data despite being trained solely with numerical data. While this supports the feasibility of training neural networks quickly and easily with simulations, it identifies the potential to use experiments for fine-tuning. 

\section{Conclusion}

This work presents a deep-learning approach to predict the angle-dependent sound absorption coefficient in a free field from a two-microphone measurement. The proposed method effectively augments a well-known experimental technique with a data-driven model that learns the edge-diffraction effect from the measurement. 
A 1D residual neural network architecture is trained to map the complex-valued transfer function between the microphones and the source elevation angle as input. Based on this, the network predicts the corresponding sound absorption coefficient as if the sample is infinite. The network is trained and validated using numerical data generated with a BEM model. It was shown that the method generalizes well numerically for different material characteristics, absorber sizes down to \SI{20}{\centi\meter}$\times$\SI{20}{\centi\meter}, and source elevation angles up to $80^\circ$. Furthermore, the method was validated experimentally in eight different scenarios, reinforcing the generalization abilities of the neural network. The network produced consistent results for varying source distances independently of variations in the measured transfer functions.

Additionally, it was shown that the network can accurately capture the influence of the wave incidence angle on the predicted sound absorption coefficients. While the predictions for the larger sample were in better accordance with the Miki model's estimation, the predictions for the smaller sample showed a more similar behavior to the impedance tube measurements. Possible reasons for the deviations between small and large samples were discussed. Although the network predictions only show minor variabilities for measurements of the same sample, a possible network sensitivity to changes in the measured transfer functions in the low-frequency range was determined. Future work should focus on further experimental validation and explore the possibility of fine-tuning the pre-trained model using actual two-microphone measurements. 



\begin{acknowledgments}
E.Z. acknowledges the financial support by the Swedish Research Council (Vetenskaprådet) under grant agreement No. 2020-04668. E.B. acknowledges the financial support by the National Research Council of Brazil (CNPq - Conselho Nacional de Desenvolvimento Científico e Tecnológico) under grant agreement No. 402633/2021-0. The authors would like to thank Saint-Gobain Ecophon (Hyllinge, Sweden) for providing the test samples, as well as Stefan Jacob, Joakim Aste, and Charlotte Aste for supporting the work involved in the experimental setups. 
\end{acknowledgments}

\section*{Author declarations}
The authors declare no conflict of interest.

\section*{Data availability}
The data that support the findings of this study are available from the corresponding author upon reasonable request.





\nocite{*}
\bibliography{References.bib}




\end{document}